\documentclass[preprint,prd,aps,tightenlines]{article}

\usepackage{authblk}

\usepackage{amsmath}
\usepackage{verbatim}
\usepackage{graphicx}
\usepackage[usenames]{color}
\usepackage{psfrag}

\usepackage{epstopdf}
\newlength{\vshift}
\input epsf.tex
\newlength{\hshift}

\usepackage{amssymb}
\usepackage{amsmath}
\usepackage{amsthm}
\usepackage{amsopn}

\def\beq{\begin{equation}}
\def\eeq{\end{equation}}
\def\bea{\setlength\arraycolsep{1.4pt}\begin{eqnarray}}
\def\eea{\end{eqnarray}}
\def\bit{\begin{itemize}}
\def\eit{\end{itemize}}

\newbox\tablebox    \newdimen\tablewidth
\def\leaderfil{\leaders\hbox to 5pt{\hss.\hss}\hfil}
\def\endtable{\tablewidth=\columnwidth 
    $$\hss\copy\tablebox\hss$$
    \vskip-\lastskip\vskip -2pt}

\def\tablenote#1 #2\par{\begingroup \parindent=0.8em
    \abovedisplayshortskip=0pt\belowdisplayshortskip=0pt
    \noindent
    $$\hss\vbox{\hsize\tablewidth \hangindent=\parindent \hangafter=1 \noindent
    \hbox to \parindent{\sup{\rm #1}\hss}\strut#2\strut\par}\hss$$
    \endgroup}
\def\doubleline{\vskip 3pt\hrule \vskip 1.5pt \hrule \vskip 5pt}

\title{Dimensionless cosmology}
\date{}

\author{Ali Narimani \thanks{anariman@phas.ubc.ca}}
\author{Adam Moss \thanks{adammoss@phas.ubc.ca}}
\author{Douglas Scott \thanks{dscott@phas.ubc.ca}}
\affil{Department of Physics \& Astronomy\\
University of British Columbia,
Vancouver, BC, V6T 1Z1  Canada}

\begin{document}
\maketitle

\abstract
{Although it is well known that any consideration of the variations of
fundamental constants should be restricted
to their dimensionless combinations, the literature on variations of the
gravitational constant $G$ is entirely dimensionful.
To illustrate applications of this to cosmology,
we explicitly give a dimensionless version of the parameters of the
standard cosmological model, and describe the physics of both Big Bang
Neucleosynthesis and recombination in a dimensionless manner. 
Rigorously determining how to talk about the model in a way which avoids
physical dimensions is a requirement for proceeding with a calculation
to constrain time-varying fundamental constants. 
The issue that appears to have been missed in many studies is that in
cosmology the strength of gravity is bound up in the 
cosmological equations, and the epoch at which we live is a crucial part
of the model. We argue that it is useful to consider the hypothetical
situation of communicating with another civilization 
(with entirely different units), comparing only dimensionless constants,
in order to decide if we live in a Universe governed by precisely the same
physical laws.  In this thought experiment, we 
would also have to compare epochs, which can be defined by giving the value
of any {\it one\/} of the evolving cosmological parameters.  By setting things
up carefully in this way one can avoid inconsistent 
results when considering variable constants, caused by effectively
fixing more than one parameter 
today.  We show examples of this effect by considering microwave background
anisotropies, being careful to maintain dimensionlessness throughout.
We present Fisher matrix calculations to estimate how well the fine structure
constants for electromagnetism and gravity can be determined with future
microwave background experiments.  We highlight how one can be misled by
simply adding $G$ to the usual cosmological parameter set. 
}

\pagebreak

\section{Introduction}

The study of the Universe at early times and on large scales might allow us to 
discover physics beyond the currently accepted 
models, namely $ \Lambda $CDM for cosmology and the standard
model of particle physics.  Since cosmology probes energies considerably 
beyond those attainable by man-made accelerators, there is hope that we 
can find evidence for new physics in this arena.
One speculative idea for extending 
our view of physics is to consider the possibility that the fundamental
constants are not actually constant.
Variation of fundamental constants, if realised, would play a key role
in theories beyond the standard model, in particular for ideas inspired by
string theory or extra dimensions.  In such theories the coupling 
constants can appear as fields which evolve in time
(see e.g.\ \cite{str1,BBN9}).
This idea is most familiar today through the many studies inspired
by claims of a redshift dependence in the fine structure constant
(e.g.\ \cite{Webb1999,Griest2010,Uzan2011,barrr}).

Since the pioneering articles of Dirac in the 1930s \cite{dirac1,dirac2}
regarding variation of the gravitational constant, $G$, 
there has been a large body of work on the possible variation of
several different constants (see e.g.~\cite{mang}).  Over time physicists 
have become more aware of the fact that only variation of {\it dimensionless\/} 
quantities can be meaningfully measured and discussed \cite{dicke,Duff2002,duf,Uzan}.
This is because any measurement in physics can be reduced to
dimensionless ratios of quantities -- in other words, a measurement
of something is always made {\it relative\/} to some other
quantity of the same dimensions.  In fact this underlies the utility of
``dimensional analysis'' \cite{Maxwell,Buckingham,Rayleigh} as a physics tool.

It is reasonable to ask if one can express
the results of any general physical observation in terms of a few dimensionless
parameters.  In most of cosmology the relevant physical 
quantities for specific problems are formed from the set 
$ P = \{ c, h, \epsilon_0 , e, G, m_{\rm p}, \\ m_{\rm e}, k, T \} $,\,
plus extensive variables, such as distance, mass, rate, redshift, etc.
Then, from dimensional analysis, the only dimensionless
quantities which can be constructed are
\begin{equation} 
\alpha_{\rm em} \equiv \dfrac{e^2}{4\pi\epsilon_0 \, \hbar \, c}, \quad
\alpha_{\rm g} \equiv \dfrac{G\, m_{\rm p}^2}{\hbar \,c}, \quad
\mu \equiv \dfrac{m_{\rm e}}{m_{\rm p}}, \quad
\theta \equiv \dfrac{k \, T }{m_{\rm p} \, c^2},
\label{ratio}
\end{equation} 
and combinations of them.
Here $\alpha_{\rm g}$ is the ``fine structure'' constant for gravity (a
notation which appears to have originated with Silk \cite{silk}), $T$ 
is the background radiation temperature,
and the other quantities are familiar physical constants \cite{alpha2}.
The importance of the dynamical variable $T$, temperature, in this set, as
opposed to the other dimensional constants, will be highlighted in the
following section. 

This analysis shows that if a specific quantity is
dimensionless (which should be the case if one is talking about 
physical measurements), and contains one of the members of $P$, 
then the other elements have to appear in a way that leads to at least one of 
the dimensionless ratios introduced in \eqref{ratio}.

There are many papers in the literature
(e.g.~\cite{Harwit1970,Harwit1971,Noerdlinger,SolheimBS,Pegg,Blake,Sumner,
Peng2005,davie,melni,Ichikawa2006,gibbon1,gibbon2,Scoccola,Sanejouand})
which try to answer a question
such as: ``What would happen to observable $X$, if the speed of light or 
the gravitational constant or Planck's constant had a
different value''.  Such questions are not well defined, since one can tune
the other parameters of the dimensional set $P$, such that the dimensionless 
ratios in \eqref{ratio} remain the same.  Since any observable is 
essentially dimensionless, 
it can only depend on these dimensionless ratios, and one cannot discuss
what happens to the observable $X$ if, for example, $G$ changes.
On the other hand one can meaningfully consider how $X$ might change 
if $ \alpha_{\rm g} $ varies. In this paper we will explore
consequences of adopting this dimensionless thinking to measuring
observables in physical cosmology.\footnote{In a somewhat different context
the importance of using dimensionless numbers in cosmology was stressed
earlier by Wesson \cite{Wesson1980}.}

As a prelude to the rest of the paper, let us
consider Big bang Nucleosynthesis (BBN) and the final abundance ratio of helium 
atoms relative to baryons -- a measurement which is clearly dimensionless.
We will discuss BBN in a dimensionless manner in more detail later on,
but it may be helpful first to explain the logic and sketch the main idea. 
When we say ``nothing happens to observable $X$ if $G$ changes'', one might 
naively think that changing $G$ would, for example, change the expansion
rate of the Universe and hence lead to observable consequences.
The point is that there is no
observable which depends solely on the expansion rate.  There also has to be 
some other rate (like the recombination rate, neutron decay rate, etc.)
that one is considering in the problem.  The ratio of these two rates 
(the expansion rate and the other relevant rate) is 
dimensionless, and therefore has to depend on the dimensionless ratios and 
not just $G$.  Based on this, the general dependence of the primordial 
helium abundance $ Y_{\rm p} = Y_{\rm p}
\left( c, h, \epsilon_0 ,e, G, m_{\rm p}, m_{\rm e} \right)$ 
can always be reduced to $ Y_{\rm p} = Y_{\rm p}
\left(\alpha_{\rm g},\alpha_{\rm em}\right)$.\footnote{And also
$\alpha_{\rm s}$ and $\alpha_{\rm w}$, as we will see later.}
Therefore, the dimensional constants should not be 
regarded as independent, and any variation in these constants should always 
be understood as a variation in the dimensionless ratios. 

In this paper we will consider different aspects of cosmology in relation 
to fundamental constants.
The next section deals with some difficulties which arise when one tries to
track a variation in the fundamental constants within the standard
framework of cosmology. 
In sections 3 and 4 we try to express the results of BBN and recombination in terms
of the dimensionless ratios defined in equation (\ref{ratio}).
We keep everything explicitly dimensionless in these sections, but also
simplify some of the equations. 
They therefore lack the full accuracy required for comparing with data, but
nevertheless demonstrate the main physics.
In section 5 we work through the publicly available codes for computing
cosmic microwave background (CMB) anisotropies, make them manifestly
dimensionless, and present the results of a variation of $\alpha_{\rm g}$ or
$\alpha_{\rm em}$ on the CMB power spectra. It is shown in this section that
variation in the gravitational constant brings a complication because of
the need to define a cosmological epoch.  We also perform a Fisher matrix 
analysis to show more explicitly how one could obtain incorrect results by 
being careless in adding constants in a dimensionful manner.
In section 6 we briefly discuss how this work could be extended to other
areas of physical cosmology.

In cosmology, Newton's constant, $G$, is particularly important.
This is because it enters into the dynamics of the background, the growth of
perturbations and large-scale geometric effects.  There have been many
studies of how one might constrain ${\dot G}/G$ using cosmological or
astrophysical data (e.g.~\cite{GBIK} and references therein),
as we will discuss further below.
It seems odd that studies of $\alpha_{\rm em}$ and $\mu$
are usually careful to point out the importance of only considering
dimensionless constants, while papers on $G$ (or even sections in review
articles) tend to ignore this entirely.  We already discussed the importance
of this issue in an earlier paper \cite{AliG}, and we found that in
general most of the published studies are unchanged if one replaces $G$ with
$\alpha_{\rm g}$.  However, as we will see, there appear to be exceptions
in some cosmological applications.  It is for this reason that we would
recommend keeping things dimensionless as much as possible.

At this point in a cosmology paper it is conventional to mention ones
choice of units and parameters.  We should point out that do {\it not\/}
set $c=1$ or $G=1$, or make any similar selection.  We also refrain from
using any particular cosmological parameters, dimensional or otherwise.

\section{Fundamental constants in cosmology}

In cosmology one requires an additional set of parameters (as well as the
dimensionless physical constants) to specify the model. 
This is a little different from setting the values of (dimensionless) 
physical constants, since the cosmological parameters are (at least 
statistically speaking) chosen from among a set of possible universes, all 
with the same physics and physical constants.  Some parameters (or ratios) may be fully
deterministic, but others could have stochastic values.  Someday we might have
a theory which tells us exactly the value for some of the parameters, but in
the current state of cosmology, we consider them as free parameters, which
are unknown a priori.  The usual set consists of (at least):
$ \{ \rho_{\rm B}, \rho_{\rm M},\rho_{\rm R}, \rho_{\rm \Lambda},
H_{\rm 0}, T_{\rm 0}, A , n, \ldots \} $.
These are the energy densities in various
components (baryons, matter, radiation and vacuum), the expansion rate,
CMB temperature, and amplitude and slope of the initial conditions.
Usually the densities are expressed in terms of the critical density (which
involves $G$) to give the set
$\{\Omega_{\rm B}, \Omega_{\rm M}, \Omega_{\rm R}, \Omega_{\rm \Lambda}\}$.
Parameters such as the amplitude $A$ and 
tilt $n$ of the scalar power spectrum probably depend
on fundamental constants, but in a way which is as yet unknown.

The obvious question that arises here is ``how do we define the cosmological
model in a dimensionless way?''  Some parameters are simple, for example the
power spectrum descriptors: we can convert from
$A$ to the dimensionless power at some fiducial
scale (e.g.\ $\Delta_{\cal R}^2(k_0)$ used by {\it WMAP\/} \cite{Komatsu},
$\sigma_8$ derived from galaxy clustering or Martin Rees' $Q^2$
\cite{TegmarkRees}); and the
slope $n$ is already dimensionless and so offers no difficulty.
The real problems come from the fact that the {\it other\/} main cosmological
parameters are epoch-dependent quantities.

We can imagine a gedanken conversation with an alien civilization 
to help sort this out. To communicate information on
the background cosmology, we would need to discuss what we
mean by ``baryons'', ``photons'', ``neutrinos'', etc., and then give their
densities in some way.  However, we still have to deal with the fact that many
of the usual parameters (the $\Omega$s, $H$, $T$, etc.) depend on time; this
is the issue which distinguished the $\theta$ parameter from the others in
\eqref{ratio}.
Hence, even if we have established that we live in the same Friedmann model
as our alien friends, we still have to determine whether we are observing that
model at the same epoch or not.

The best way to do this would be to agree on a fiducial period in the
evolution of the Universe and then discuss where we are relative to that.
An obvious
choice is the epoch of matter-radiation equality (discussed in a related
context in \cite{TegmarkARW}), but there are plenty
of other possibilities: when the Hubble rate is equal to a particular
reaction rate; when the Thomson optical depth is unity; when matter has the
same energy density as the vacuum, etc.  Assuming that the Universe has
flat geometry (a clearly dimensionless statement), then we can give the values
of the $\Omega$s at the agreed fiducial epoch, together with one number
to fix that epoch, say the value of $\theta$ at equality.
Then we only need to give
the value of {\it one\/} parameter today in order to fix the epoch at which
we live (making sure this is a dimensionless parameter of course).  This could
be any one of the $\Omega$s today, or the value of $z_{\rm eq}$, or the value
of $H_0 t_0$, or $\theta_0$ ($\equiv kT_0/m_{\rm p}c^2$).
Anything which is changing essentially monotonically in time, and is
dimensionless, will do (so $H_0t_0$ is fine now, but useless billions of
years ago, when it was hardly changing).

This gives rise to two complications when describing the observable
Universe, which are not there when one is discussing models
of laboratory physics.
Firstly, if one is not careful, then it is possible to effectively fix
{\it more than one\/} of the parameters today,
leading to inconsistent results.  The 
second part is that different choices of the ``what-time-is-it?''\
parameter are not entirely equivalent, because some contain a dependency on
other physics parameters. We will show effects of this in section $5$.

Take for example the choice of either the CMB temperature $T_0$, or Hubble
constant, $H_0$ (made suitably dimensionless, using the Compton coupling
time, or some other timescale involving atomic physics).  These two are related 
to each other via $\Omega_{\rm R}$ and $\alpha_{\rm g}$ using the Friedmann
equation
\beq 
\label{epoch}
{H_0^2=\dfrac{8 \pi^3}{45} \, \dfrac{\alpha_{\rm g}}{\Omega_{\rm R}\hbar^2} \, 
\dfrac{k^4 \, T_0^4}{m_{\rm p}^2 \, c^4}\,.
}\eeq
Therefore, either $H_0$ or $T_0$ is enough to determine today's epoch,
and it is not consistent to use the Friedmann equation and treat {\em both}
of these as free parameters.
However, it turns out that it matters whether we choose $T_0$ or $H_0$, 
since the pair $ \left(T_0,H_0 \right) $ in {\em our} Universe would not
satisfy the above equation in a universe with a different $\alpha_{\rm g}$.
We return to this issue in section 5.

\section{Dimensionless BBN}

Big Bang nucleosynthesis (BBN) is an area of astrophysics which has been
thoroughly investigated for signs of variation in the fundamental constants 
(see e.g.~\cite{str1,nol,landau} for non-gravitational couplings
and BBN).  Among the constants which are effective during BBN, the 
gravitational constant plays a key role, and hence there have been 
many published studies of the effects of a varying $G$
on the primordial abundance of the light elements 
\cite{Dent,Li,richard,copi,wet,Bambi,fabris,orlov,irvine,Yang}. However, since 
$G$ is a dimensional constant, we calculate the abundance of helium synthesized
during BBN in an explicit dimensionless way in this section, focusing on the 
dominant parts of the physics and neglecting some of the finer details.  Our 
calculation, though crude, shows the role of 
$ \alpha_{\rm g} $ in primordial nucleosynthesis and explicitly reveals 
the dimensionality in the relevant physics.

The key parameter in BBN is the ratio of number densities of neutrons
to protons, which can be defined as:
\beq{
R\equiv\dfrac{n_{\rm n}}{n_{\rm p}}=e^{\tiny{-u}}.
}\eeq
The quantity $u$ is the ratio of mass difference to freeze-out 
temperature:
\begin{equation}
u \equiv \dfrac{\left(m_{\rm n}-m_{\rm p}\right)c^2}{k\, T_{\rm f}}.
\end{equation}
$ T_{\rm f} $ is explicitly the temperature at which the following
reactions freeze out:
\begin{equation}
{{\rm e}^- + {\rm p} \to \nu_{\rm e} + {\rm n}; \qquad
{\rm p} + \bar{\nu}_{\rm e} \to {\rm e}^+ + {\rm n} }.
\end{equation}
According to Bernstein \cite{Bern}, the rate for these reactions is:
\beq{
\lambda=\dfrac{255}{\tau_{\rm n} \, u^5}\left(12+6u+u^2\right),
}\eeq
where $ \tau_{\rm n} $ is the lifetime of the neutron.
In order to identify the effects of the gravitational coupling constant,
we can write this lifetime as
\beq{
\tau_{\rm n} = \dfrac{f_1(\alpha_{\rm w}, m_{\rm e}/m_{\rm p}) h}{m_{\rm p} \,{c^2}},
}\eeq
where $f_1$ is a dimensionless function of the weak coupling constant,
$\alpha_{\rm w}$, electron to proton mass ratio, $\mu$,
and possibly some other mass ratios and 
constants, but not a function of $\alpha_{\rm g}$.
The freeze-out temperature is set by the 
equality of this rate and the expansion rate of the Universe:
\beq
\lambda = H \quad \Rightarrow \quad
w \, u^3 -u^2 -6u-12=0,
\eeq
where we have defined $ w $ through
\begin{center}
$ w \equiv \dfrac{f_1}{255} \sqrt{\dfrac{8 \pi g_*}{3}}
\, \left(1-\dfrac{m_{\rm p}}{m_{\rm n}} \right)^2 
\, \alpha_{\rm g}^{1/2} $,
\end{center}
and $ g_* $ takes care of the number of relativistic species.
This cubic equation can be solved to obtain the 
freeze-out temperature as a function of gravitational coupling:
\begin{equation}
u=\dfrac{1}{3} \, \dfrac{\left(18 w+1\right)^{2/3}
+\left(18 w+1\right)^{-1/3} +1}{w}.
\end{equation}

After this time the most important remaining reaction is the $\beta$-decay of
neutrons, which continues until the formation of
deuterium.  Deuterium formation is delayed due to the large
photon-to-baryon ratio.  Its abundance is fixed 
at a temperature $ T_{\rm D} $, which is roughly given by the
equality of the number density of photons with the number density of the
deuterium which has been formed:
\beq{
\exp \left( \dfrac{B_{\rm D}}{k \, T_{\rm D}} \right) \eta = 1.
}\eeq
Here
$ \eta $ is the ratio of baryons to photons and $ B_{\rm D} $ is the
binding energy of deuterium, which can be written as a 
dimensionless function of the strong and electromagnetic fine structure
constants times the proton mass, 
$ B_{\rm D} = f_2 (\alpha_{\rm s},\alpha_{\rm em}) \, m_{\rm p} \, c^2 $.
The function $f_2$, like $f_1$, is dimensionless and does not depend on
$\alpha_{\rm g}$.
The temperature $ T_{\rm D} $ could also be converted to an age
through the Friedmann equation.  By the time of deuterium formation 
neutrinos have already frozen out and electron-positron pairs have
annihilated.  Thus the ratio of the age of the Universe to the neutron
lifetime becomes
\beq{
v \equiv \dfrac{t}{\tau_{\rm n} } =
\dfrac{1}{f_1 \, f_2^2 \, \ln ( \eta) ^2} \, \sqrt{\dfrac{45}{32 \, \pi ^3 }}
\, \alpha_{\rm g}^ {-1/2},
}\eeq
and the primordial helium fraction (by mass) is
\beq{
Y_{\rm p} = 2 \, \dfrac{R}{1+R} e^{-v}.
}\eeq
We have aimed at an expression for the helium abundance which is manifestly
dimensionless, i.e.\ it only depends on dimensionless ratios of physical 
quantities, including $ \alpha_{\rm g} $.
This simple analysis leads to a primordial helium fraction of about
$ Y_{\rm p} \sim 0.22 $, which is within $ 10 $ percent of the value
coming from more complicated numerical BBN codes.
Now we can put all this together to track the effects of a possible
variation of $ \alpha_{\rm g} $
on $ Y_{\rm p} $:
\begin{eqnarray}
\delta Y_{\rm p} &=& \dfrac{e^{- \left( u+v \right) } }{\left(1+R\right)^2 }
\, \alpha_{\rm g}^{{1/2}} 
 \left[ \dfrac{u^3}{3 \, w \, u^2 - 2u -6 } \dfrac{f_1}{255}
\sqrt{\dfrac{8 \, \pi g_*}{3} }
\left(1-\dfrac{m_{\rm p}}{m_{\rm n}} \right)^2 \right. \nonumber \\
&& \left. \! +\sqrt{\dfrac{45}{32 \, \pi ^3 }} 
\dfrac{1+R}{f_1 \, f_2^2 \, \ln ( \eta) ^2} \, 
\alpha_{\rm g}^{-1} \right] \dfrac{\delta \alpha_{\rm g}}{\alpha_{\rm g}}.
\end{eqnarray}
The first term in the square brackets comes from the change in the neutron's
freeze-out
fraction and the second is due to a change in the age of the Universe.
Both terms have the same order of magnitude $ \left( \sim 10^{-2} \right) $
and have the same sign.  A higher $ \alpha_{\rm g} $ leads to more neutrons
at the freeze-out time and a lower age for the Universe, 
both of which enhance the primordial helium fraction.

If one takes the measurement error on $ Y_{\rm p} $ to be $\sim0.005$,
and assumes a power-law variation with time of the form
$ \alpha_{\rm g}\propto t^{-x} $,
this simple analysis shows that $x$ should be less than $0.005$, which is
consistent with the results of other studies (see e.g.~\cite{Yang}).

We end this section with a discussion of other published studies of BBN.
Despite this literature being almost entirely dimensionful, we find that
most of the 
papers on variation of $G$ within BBN can be considered to be valid
if one simply reinterprets a
variation in $G$ as a variation in $\alpha_{\rm g}$. However, this is not
always the case, particularly where extra physics is considered.
Sometimes statements are made like ``whenever you see
$G$, interpret it as $G$ times some quark mass $m_{\rm x}$, or a combination
of $G$ and $ \Lambda_{\rm QCD}$''
(see e.g.~\cite{mang}).  But actual measurements of $G$ have always been made
using normal atoms, and therefore when using physical equations to
derive some proposed time variation for $G$ (or more
properly, $ \alpha_{\rm g} $), then $ \alpha_{\rm g} $ is effectively
the one introduced in equation \eqref{ratio}.  In this sense,
those papers which are assuming a simultaneous time variation for $G$ and
$m_{\rm p}$, are not valid (see e.g.~\cite{Dent,mang}).
  
\section{Redshift of recombination} 

Cosmological recombination of hydrogen is mainly controlled by the
population of the first excited state \cite{zel}.  This is because 
of the high optical depth for photons coming from transitions direct to 
the ground state.  The rate of recombination to this first excited state 
is given as \cite{Spitzer,dodel}
\beq{
\Gamma = n_{\rm B} \, \alpha^{(2)} = 9.78 \, n_{\rm B}
\left( \dfrac{E_0}{k \, T} \right) ^ {1/2} \,
\ln \left( \dfrac{E_0}{k \, T} \right) \,
\dfrac{\hbar^2 \alpha_{\rm em}^2}{m_{\rm e}^2 \, c}.
}\eeq
Here ``${\rm B} $'' stands for baryons (i.e.\ protons and 
neutrons), ``${\rm e}$'' for electrons, 
$ \alpha^{(2)} $ is the recombination rate to the second energy 
level of hydrogen and $ E_0 $ is the binding energy of hydrogen,
which is equal to $\alpha_{\rm em}^2 m_{\rm e} c^2/2$.
We have assumed that all of the atoms are ionized.
It also simplifies things considerably (without qualitatively changing the
physics) if we ignore the mass difference of protons and neutrons, i.e.\ set
$m_{\rm B} =m_{\rm p} $, so that
\begin{equation}
n_{\rm B}= \dfrac{\rho_{\rm B}}{m_{\rm B}} =
\dfrac{\rho_{\rm 0B}}{m_{\rm p}} \left( \dfrac{T}{T_0} \right) ^3.
\end{equation}
This gives a rate
\beq{
\Gamma = 7.0 \left( \dfrac{m_{\rm e} \, c^2 }{k \, T_0} \right)^{1/2}
\dfrac{\hbar^2 \alpha_{\rm e}^3}{m_{\rm e}^2 \, c} \,
\dfrac{\rho_{\rm 0B}}{m_{\rm p}} \,
\ln \left( \dfrac{E_0}{k \, T} \right) \,
\left( \dfrac{T}{T_0} \right)^{5/2}.
}\eeq
Assuming for further simplicity that
$\Omega_{\rm M} =1 $ (this assumption could easily be relaxed later), 
the Hubble constant is
\beq{
H = \left( \dfrac{8 \pi G \rho_0}{3} \right) ^ {1/2} \,
\left( \dfrac{T}{T_0} \right) ^ {3/2},
}\eeq
and one can also use the equalities
\begin{equation}
\rho_{\rm 0B} = \rho_{\rm 0R} \dfrac{\Omega_{\rm B}}{\Omega_{\rm R}}, \quad
\rho_{0} = \rho_{\rm 0R} \dfrac{\Omega_{\rm M}}{\Omega_{\rm R}},
\end{equation}
along with
\beq{
\rho_{0\rm R} = aT_0^4 = {8\pi^5 k^4 \over 15 c^3 h^3}T_0^4.
}\eeq
Putting all of these together, one can find an expression for the redshift of
recombination:
\beq
\label{recom}{
1+z_{\rm r} \propto \dfrac{T_{\rm r}}{T_0}
\propto \alpha_{\rm g}^{1/2} \, \alpha_{\rm em}^{-3} \,
 \mu^{3/2} \, \theta_0^{-3/2} \,
\left( \dfrac{\Omega_{\rm M}}{\Omega_{\rm R}} \right)^{1/2} \,
\left( \dfrac{\Omega_{\rm B}}{\Omega_{\rm R}} \right)^{-1}.
}\eeq

Again we can see that an observable, which is the redshift of
recombination, depends only on the dimensionless ratios of physical quantities, 
together with some other dimensionless parameters.
Although this expression contains much of the essential physics,
unfortunately the numerical factor (which we neglected to write down) is
far from correct.  That is because we have not taken account of the
partial hydrogen ionization.  It would be possible to take this further
by using an approximation to the Saha equation to correct for the ionized
fraction, which would change the scalings with the dimensionless parameters.
However, this rapidly gets complicated, and for accuracy one really needs the
numerical solution to the relevant differential equations (which we will
do in the next section).

Nevertheless, the main point is that
one can see an important dependence
on the dimensionless quantities of interest.
The change in the redshift of recombination is one of the main cosmological
effects of varying $\alpha_{\rm g}$.  There are also other effects on
CMB anisotropies, as we will see in the next section.

\section{Varying constants and CMB anisotropies} \label{sec:cmb}

So far we have written down expressions which were explicitly 
dimensionless, but at the expense of accuracy.  Let us now 
numerically explore the effects of a varying $ \alpha_{\rm em} $ or 
$ \alpha_{\rm g} $ on the CMB anisotropies.  We use {\sc recfast} v$ 1.5
$ \cite{douglas} for the recombination history of the Universe and
{\sc cmbfast} \cite{cmb} for the calculation of CMB angular power spectra.
It turns out that one would face some difficulties if one were to
blindly dive into the codes 
and try to change ``$G$'' or add some factors of $\alpha_{\rm em}$ in the
relevant parts.  There are several issues which should be considered
before one can promote these constants into dynamical parameters in the code.
It turns out that 
this choice not only affects the feasibility of the problem, but also the 
outcome of the numerical calculation.  As an example, {\sc cmbfast} is written
so that it uses the total density contributions, 
$ \{ \Omega \} $, both for the strength of inertia in acoustic modes, and 
as gravitational charge.  Therefore, if one chooses these ratios as the free
parameters, in place of densities, it is much easier to trace the effects of a
variation in the gravitational constant, because there is an additional factor
of $G$ wherever the code needs the $\Omega$s as seeds for gravitational 
collapse.

\subsection{Recombination through RECFAST}
\begin{figure}[t]
\includegraphics[totalheight=0.4\textheight ,
width=1\textwidth ,viewport=0 0 380 280]{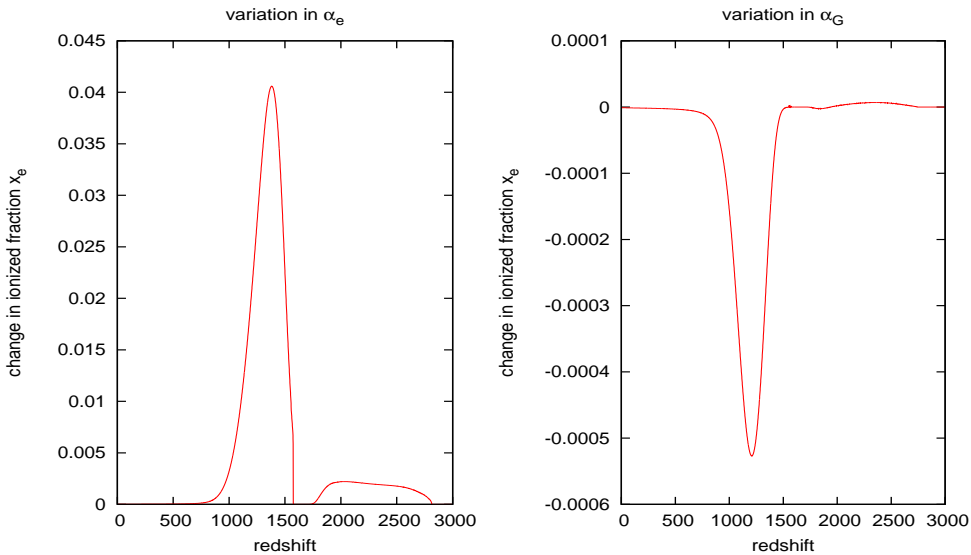}
\caption{\label{fig:rec} Effects on the ionization history of 1\% increases
in $\alpha_{\rm em}$ (left) and $\alpha_{\rm g}$ (right).}
\end{figure}

One can trace the effects of a varying fine structure constant
or gravitational constant on the process of recombination through 
{\sc recfast}
(see e.g.~\cite{harari,hanns,kaplin,Battye2001}).  In order to convert the code
to a form where it can run with a different $ \alpha_{\rm em} $, one 
has to track all of the relevant dependencies~\cite{adam}. 
Most of this shows up in the 
energy levels, since $ E \propto \alpha_{\rm em}^2 $.
The ``Case-B'' recombination rate (see Rybiki and Lightman p.\,282 \cite{ryb})
has an $ \alpha_{\rm em}^5 $ dependence and the complete analytic form of the 
2s--1s two photon rate (\cite{chuba, spt}) varies as 
$ \alpha_{\rm em}^8 $.  Triplet transitions of helium (discussed 
in \cite{grz}), case-B recombination and two photon transition
rates for helium have the same scaling as for hydrogen.  With this 
information in hand one can trace the effects of a different 
$ \alpha_{\rm em} $ (or a time-dependent $ \alpha_{\rm em} $) on recombination.
The $G$ dependence for recombination lies entirely in the Hubble 
constant.  Putting the rates together and defining the
dimensionless ratios,
\beq{
A \equiv \dfrac{n_{\rm p} \, \alpha }{H}, \qquad
B \equiv K \, \Lambda n_{\rm p}, \qquad
C \equiv K \, \beta \, n_{\rm p}, \qquad
D \equiv \dfrac{\beta}{H},
}\eeq
the Saha equation \cite{douglas} will take the form
\beq{
\dfrac{dx_{\rm p}}{dz} = \left[x_{\rm e} \, x_{\rm p} \, A
- \left(1-x_{\rm p}\right)D \right]
\dfrac{\left[1+B (1-x_{\rm p} ) \right] }
{(1+z) \left[1+(B+C) (1-x_{\rm p}) \right] }.
}\eeq
This is explicitly for hydrogen, but there is a similar form for helium too.
These dimensionless ratios have the following dependencies on the coupling
constants:
\begin{eqnarray}
&&A \left(\alpha_{\rm em} , \alpha_{\rm g} \right) 
\propto \alpha_{\rm em}^5 \, \alpha_{\rm g}^{-0.5}\, ; \\&&
B \left(\alpha_{\rm em} , \alpha_{\rm g} \right) 
\propto \alpha_{\rm em}^2 \, \alpha_{\rm g}^{-0.5}\, ; \\&&
C \left(\alpha_{\rm em} , \alpha_{\rm g} \right) 
\propto \alpha_{\rm em}^{-1} \, \alpha_{\rm g}^{-0.5}\, ; \\&&
D \left(\alpha_{\rm em} , \alpha_{\rm g} \right) 
\propto \alpha_{\rm em}^5 \, \alpha_{\rm g}^{-0.5}\, .
\end{eqnarray}
Figure~\ref{fig:rec} shows the results of a $1\%$ increase in the fine 
structure constant and separately of
the gravitational constant through the history of 
recombination.  This is consistent in sign with equation \eqref{recom} -- an
increase in $\alpha_{\rm em}$ leads to a higher ionization fraction, which
leads to a lower redshift for recombination.  It is also 
noticeable that even a 1\% increase in $\alpha_{\rm em}$ can lead to more
than 1\% variation in the recombination history, while such a variation
in $\alpha_{\rm g}$ does not leave any significant trace, due to the
weaker power-law scalings above and the relatively thin 
last scattering surface.

\subsection{Perturbations through CMBFAST}

The effects of an alteration in the recombination history could in
principle be measured through the CMB power spectra.  This section complements
section~3 in terms of comparing a
hypothetical variation in the fundamental constants with known physics
at a particular epoch, $z \sim 1000 $ in this case,  rather than
$z \sim 10^9$ for BBN.  As is discussed in \cite{adam}, a different
fine structure constant leaves its main imprint on the recombination
history, and this shows up in {\sc cmbfast} through two main effects:
$(1)$ the derivative of the opacity, which is proportional to Thomson
scattering, with an $ \alpha_{\rm em}^2 $ dependence; and $(2)$ the number
density of free electrons, which is basically the ``freeze-out'' value at the
end of recombination.  Therefore, one can almost ignore the effects of a
different fine structure constant after recombination.

\begin{figure}[t]
\includegraphics[totalheight=0.4\textheight ,
width=1\textwidth ,viewport=0 0 380 280]{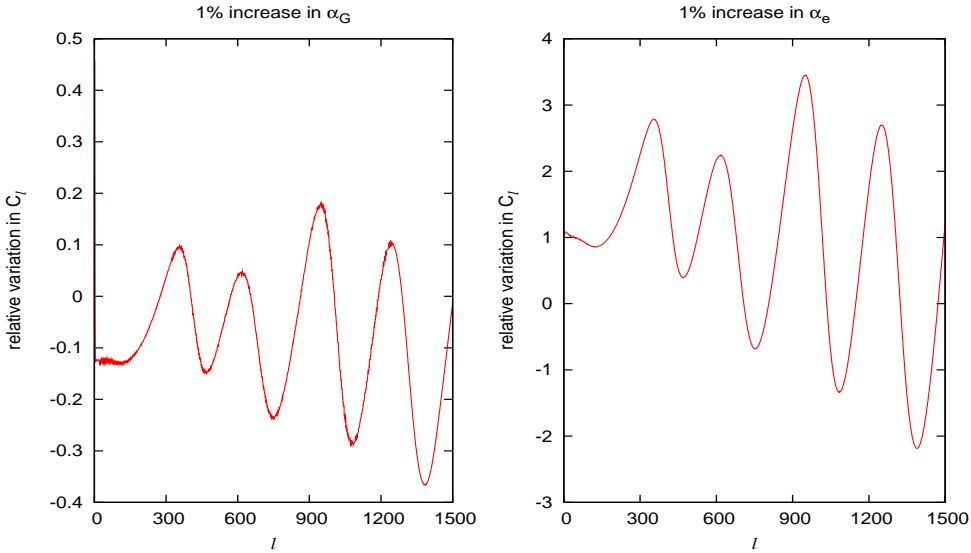}
\caption{\label{fig:cmb1} Effects on the CMB anisotropies of a 1\%
increase in $\alpha_{\rm g}$ (left) and $\alpha_{\rm em}$ (right).
The vertical axis is percentage.}
\end{figure}

The case is different
for $ \alpha_{\rm g} $.  As is discussed in \cite{Uzan2}, a different $G$ will
lead to a different sound horizon at the last scattering surface and a
different distance from this surface to us, the combination of which will
change the position of the peaks of the power spectrum.  There is also an
integrated Sachs-Wolfe effect (ISW) which can enhance the amplitude of
perturbations.  This is examined in \cite{zahn} and, as explained in
\cite{chan} for the case of a constant but different $G$, the effects are
much less important than they are for a time-varying $G$.
This is partly because there is no ISW effect, but also because the changes
on the sound horizon and distance to last scattering partially cancel.
The results of a $1 \% $ increase
in $\alpha_{\rm em}$ and $\alpha_{\rm g}$ are shown in Fig.~\ref{fig:cmb1}.

In \cite{zahn}, the authors propose a scaling in $G$,
($G\rightarrow \lambda^2 \, G $), which appears equivalent to the scaling of
$\alpha_{\rm g}$ considered here.  However, we cannot confirm their results, since they have assumed that
$ \{ T_{\rm 0}, H_{\rm 0}, \Omega_{\rm R} \} $ have the {\em same} values as
in the standard model.  This is incorrect, because these
three are not independent and a change in $\alpha_{\rm g}$ will result in a
different value for at least one of them (depending on which one 
we have chosen as our dependent variable).

\begin{figure}[t]
\includegraphics[totalheight=0.4\textheight ,
width=1\textwidth ,viewport=0 0 380 280]{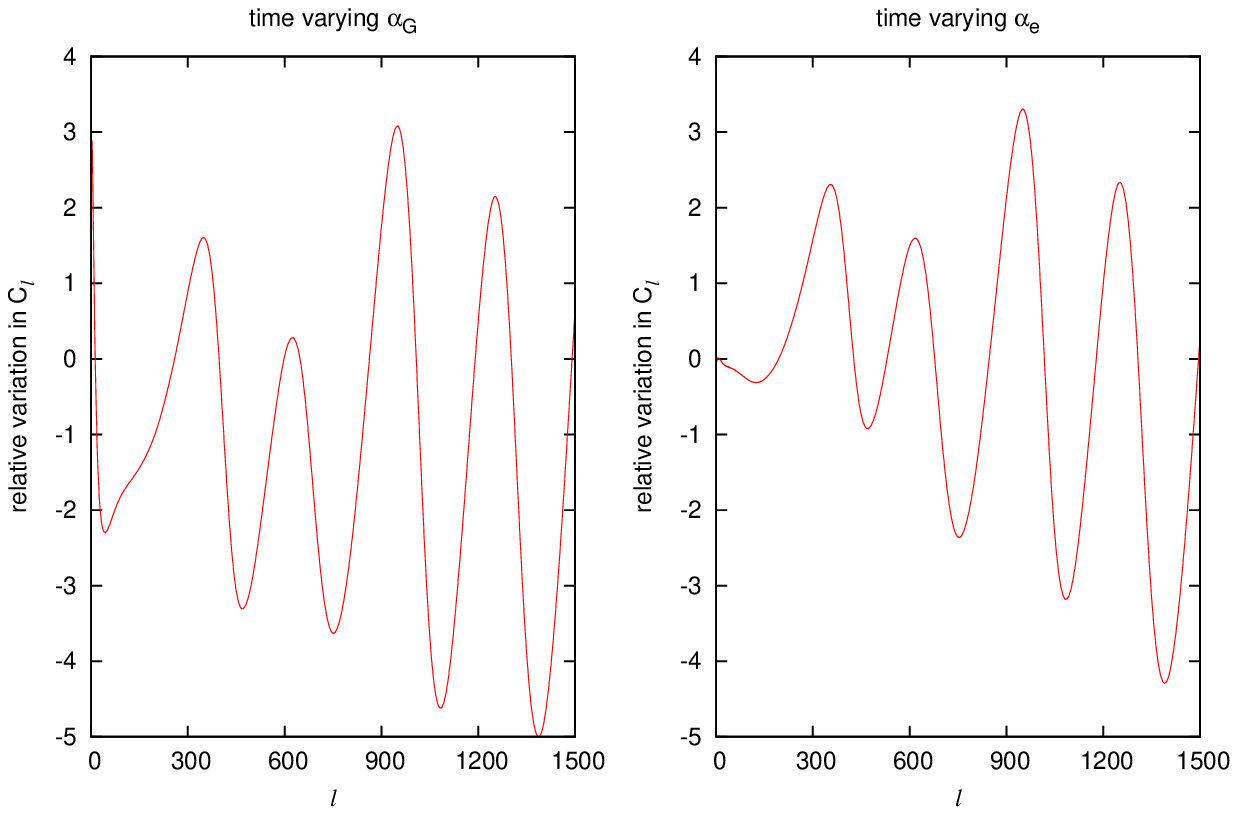}
\caption{\label{fig:cmb2} Effects on the CMB anisotropies of time-varying
constants.  We have specifically used $\alpha_{\rm g} \propto t^{-0.002}  $ 
(left) and $\alpha_{\rm em} \propto t^{-0.002}$ (right). 
The vertical axis is percentage. }
\end{figure}

An example of time varying $ \alpha_{\rm g} $ or $\alpha_{\rm em}$ results are
shown in Fig.~\ref{fig:cmb2}.  We have explicitly used
$\alpha\propto t^{-0.0002}$.  It is easy to instead imagine power-law
variations in redshift, conformal time or some other variable.
Specific ideas for time-varying constants may lead to specific forms for the
function of time, e.g.\ $\alpha_{\rm em}(t)$.  Following such detailed
predictions are beyond the scope of the present paper.  However, the 
methodology set out here should be useful in testing any explicit model
that is proposed.

Our study is certainly not the first to explore the effects of variable
fundamental constants on the CMB (see e.g.~\cite{zahn,galli,chan}
for $G$ and \cite{Avelino,martins,Rocha,hann,Martins2010}
for $ \alpha_{\rm em} $).  The main point  
here is that we have insisted on doing everything in a dimensionless way, to
avoid finding any apparent effects which are simply a result of the choice
of units. 
As an example, \cite{journal} suggests that one can search for a
violation of the Strong Equivalence Principle using the CMB power spectra.
The method, though at first sight quite elegant, is totally dimensional.
When one starts to make it dimensionless,
it becomes evident that the whole argument has to be reconsidered.
A basic assumption is that a variation in $G$ can make a difference in
gravitational mass relative to the inertial mass of baryons.  However, this
is in general wrong, since 
one can \textit{always\/} use the freedom of setting the 
gravitational charge and inertial mass of a given body (and only for that one
object) equal to each other.

Several particular ideas for exploring the variation of $G$ within
cosmology have been described elsewhere.  Although we have not comprehensively
checked all previous studies, we are suspicious of any
in which the dimensionful quantity $G$ is discussed along with other
dimensionful constants, such as $H_0$ (e.g.\
\cite{Umezu2005,Galli2009,Luetal2010,Landau2010,Mukhopadhyay}).

\subsection{Fisher matrix analysis}
\label{sec:fisher}
\begin{table*}
\nointerlineskip
\setbox\tablebox=\vbox{
\newdimen\digitwidth 
\setbox0=\hbox{\rm 0} 
\digitwidth=\wd0 
\catcode`*=\active 
\def*{\kern\digitwidth} 
\newdimen\signwidth 
\setbox0=\hbox{{\rm +}} 
\signwidth=\wd0 
\catcode`!=\active 
\def!{\kern\signwidth} 
\newdimen\pointwidth 
\setbox0=\hbox{\rm .} 
\pointwidth=\wd0 
\catcode`?=\active 
\def?{\kern\pointwidth} 
\halign{\hbox to 1.75in{#\leaderfil}\tabskip=2.0em&
\hfil#\hfil\tabskip=1.0em&
\hfil#\hfil\tabskip=1.0em\cr
\noalign{\doubleline}
\omit \hfil Parameter\hfil& Value& Uncertainty\cr
\noalign{\vskip 3pt\hrule\vskip 4pt}
\omit \hfil Coupling constants\hfil & \omit& \omit\cr
\noalign{\vskip -10pt}
\omit \hrulefill& \omit& \omit\cr
$\alpha_{\rm em}$& \omit$7.297\times10^{-3*}$& $\ldots$\cr
$\alpha_{\rm g}$& \omit$5.906\times10^{-39}$& $\ldots$\cr
\noalign{\vskip 3pt\hrule\vskip 4pt}
\omit \hfil Quantities at equality\hfil& \omit& \omit\cr
\noalign{\vskip -10pt}
\omit \hrulefill& \omit& \omit\cr
$\Omega_{\gamma}$& 0.296& $\ldots$\cr
$\Omega_{\nu }$&    0.204& $\ldots$\cr
$\Omega_{\rm CDM}$&0.415& 0.004\cr
$\Omega_{\rm B}$&  0.085& 0.004\cr
$\Omega_{\rm \Lambda}$& \omit$4.4**\times10^{-11}$& \omit$1.2\times10^{-11}$\cr
$\theta$&\omit$8.0**\times10^{-10}$& \omit$0.3\times10^{-10}$\cr
\noalign{\vskip 3pt\hrule\vskip 4pt}
\omit \hfil Definitions of ``now''\hfil& \omit& \omit\cr
\noalign{\vskip -10pt}
\omit \hrulefill& \omit& \omit\cr
$z_{\rm eq}$& 3200& 130\cr
$\theta_0$& \omit $2.503\times10^{-13}$& $\ldots$\cr
\noalign{\vskip 3pt\hrule\vskip 4pt}
}}
\endtable
\caption{Dimensionless cosmological model parameters.  We have assumed massless
neutrinos here, contributing to the ``radiation'' by the usual factor of 0.68
times the photon energy density.  We have also assumed a flat background,
and calculated values and uncertainties using the Markov chains from
the {\it WMAP\/} 6-parameter fits \cite{Komatsu}.}
\label{tab:dimensionless}
\end{table*}

Let us be more explicit in defining the cosmological model which describes
our Universe.
In terms of the standard cosmological picture, one could define dimensionless
parameters at some fiducial epoch.  If we choose matter-radiation equality
as this epoch, then we can define the background cosmology by giving the
values of the $\Omega$s {\it then}, as shown in
Table~\ref{tab:dimensionless}.  We also need to give a quantity to set the
epoch at equality, and we choose $\theta$.  Finally we need to provide one
parameter (out of a wide range of possibilities) to define the time
{\it today}.  We have pointed out how it is possible to reach invalid
conclusions about the variation of fundamental constants (particularly $G$)
by fixing too many parameters in the cosmological model.  In this section
we show this in practice through a Fisher matrix calculation to forecast
parameter uncertainties for future observations. 

The details of Fisher matrix analysis 
for CMB anisotropies are fully explained in \cite{tegmark}.
Here we have used the characteristic 
values for the {\it Planck\/} satellite \cite{Planck}, and summed over the
$\{ TT, TE, EE \}$ channels, where $T$ stands for temperature and
$E$ is the E-mode polarization.

Specifying the value of $\Omega_{\rm \Lambda}$ at equality and
the ratio $X (\equiv \Omega_{\rm CDM}/\Omega_{\rm B})$ suffices to give
the correct fraction of the energy density in 
each sector at equality, i.e.\
$\{ \Omega^{\rm eq}_{\rm CDM},\Omega^{\rm eq}_{\rm B},
 \Omega^{\rm eq}_{\nu}, \Omega^{\rm eq}_{\gamma},
 \Omega^{\rm eq}_{\rm \Lambda} \}$, in a flat Universe.  Therefore, the
equality epoch is completely defined by the set of parameters
$\{\Omega^{\rm eq}_{\rm \Lambda},X,\theta_{\rm eq} \}$.
And the description of the Universe's observables
will be complete by providing the time today, $T_0$ (or $\theta_{\rm 0}$).

The publicly available CMB anisotropy codes are written in terms of the
parameters today, rather than at a physical epoch, such as
equality.   To use these codes with variable $G$ it is important to first
ensure that the simpler situation is being considered properly,
in which we have one value of $G$ at recombination and another value today.
This is done by relating the values of the parameters at equality
with those defined today.  The easiest of the set to treat is $X$,
which is constant if neither the dark matter particles or the baryons
can transform into anything else after equality.  The temperature
$T_{\rm eq}$ is $T_0 \Omega_{\rm M}/\Omega_{\rm R}$,
where $\Omega_{\rm M}=\Omega_{\rm CDM} +\Omega_{\rm B}$
and $\Omega_{\rm R}=\Omega_\gamma+\Omega_\nu$.
The dark energy at equality is
\beq
{
 \Omega^{\rm eq}_{\rm \Lambda}
 = \dfrac{\Omega_{\rm \Lambda}}{\Omega_{\rm \Lambda}+
  \Omega_{\rm M}\left(1+\dfrac{\Omega_{\rm M}}{\Omega_{\rm R}}\right)^3+
  \Omega_{\rm R}\left(1+\dfrac{\Omega_{\rm M}}{\Omega_{\rm R}}\right)^4}.
}
\eeq

The set of parameters chosen for the Fisher matrix calculation are:
$\{X,\theta_{\rm eq},\theta_0, \tau,A_{\rm s},n,\alpha_{\rm g},\alpha_{\rm e}
 \}$.  Here, $\tau$ is the optical depth of the 
Universe after reionization, $A_{\rm s}$ is the amplitude of the power spectrum
of scalar perturbations (which could easily be made dimensionless) and $n$ is
its power-law slope. 
One could also perform the same calculation in a simplistic and incorrect
way by choosing the wrong set of parameters, $\{h_0, \Omega_{\rm CDM},
\Omega_{\rm B},\tau,A_{\rm s},n,\alpha_{\rm g},\alpha_{\rm e} \}$, that are
both time dependent and inter-dependent on each other, and by
over-constraining the observation epoch, i.e.\ ignoring Equation~\ref{epoch}.
In order to perform a fully consistent analysis, one should also take
care to normalize the power spectra to the initial conditions and not to
the large angle CMB anisotropies today; {\sc CAMB}\cite{CAMB} was chosen 
for the Fisher matrix analysis, since one can choose either option
for normalization.

\begin{figure}[t]
\includegraphics[totalheight=0.4\textheight ,
width=0.7\textwidth ,viewport=0 0 380 280]{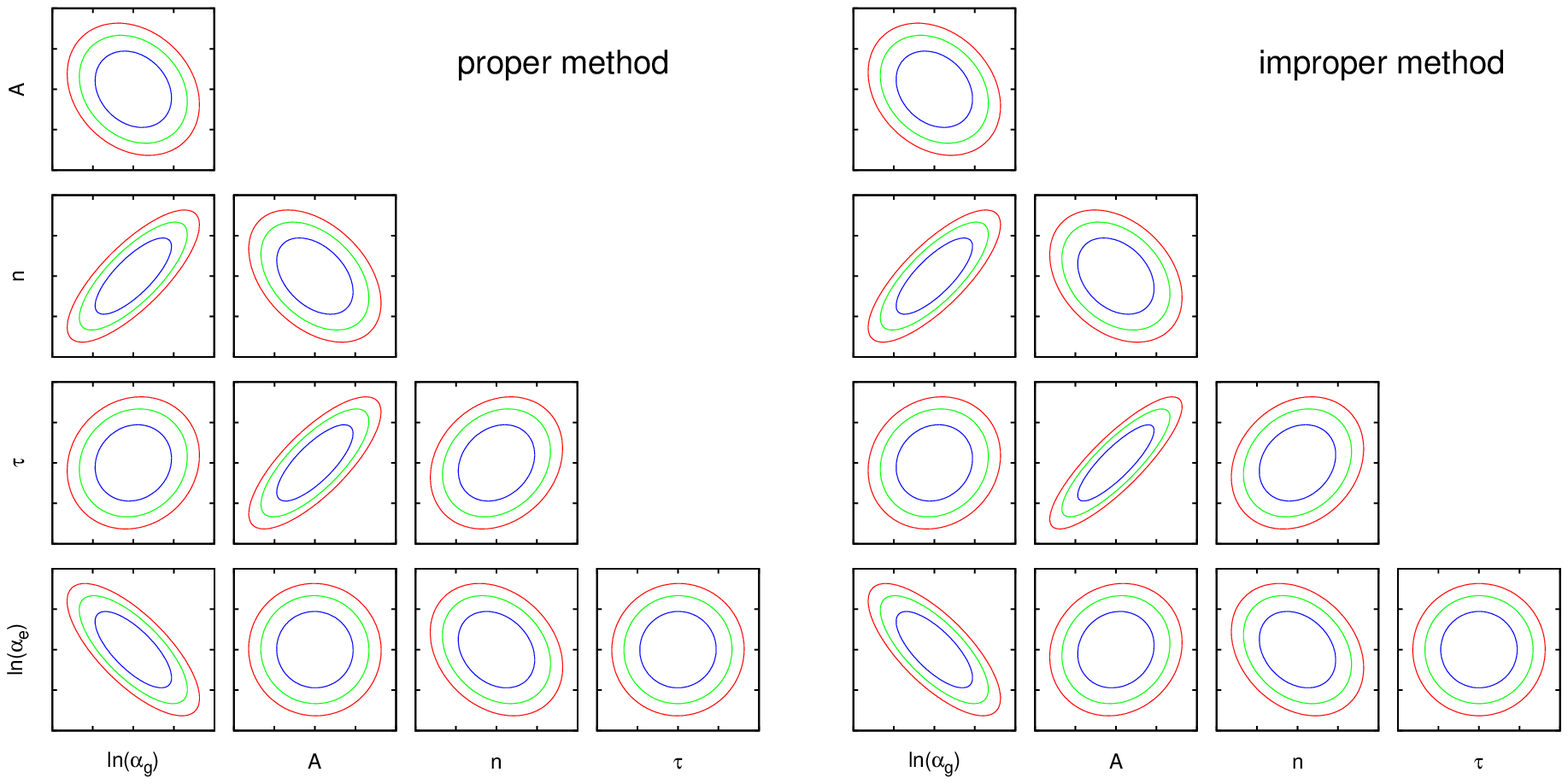}
\caption{\label{fig:correlation}Comparison of the correlation among selected
parameters for the two different methods described in Section~\ref{sec:fisher} 
for the Fisher matrix analysis.  The diagram on the left corresponds to
the correct method, while the panels on the right show the results of the
naive (and incorrect) approach.
 }
\end{figure}

The $1\sigma$ error bars for the two different methods are contrasted in
Table~\ref{tab:fisher} and the correlations among the parameters are
shown in Fig.~\ref{fig:correlation}.
The conclusion is that the error bar on $\alpha_{\rm g}$ changes by
approximately a factor of $2$ between the two different approaches,
and $\alpha_{\rm g}$ shows a slightly stronger 
correlation with other parameters in the correct approach compared to the
incorrect one.  This underscores the need to be careful when considering
the effect of variable constants on cosmological observables.

\begin{table*}
\nointerlineskip
\setbox\tablebox=\vbox{
\newdimen\digitwidth 
\setbox0=\hbox{\rm 0} 
\digitwidth=\wd0 
\catcode`@=\active 
\def@{\kern\digitwidth} 
\newdimen\signwidth 
\setbox0=\hbox{{\rm +}} 
\signwidth=\wd0 
\catcode`!=\active 
\def!{\kern\signwidth} 
\newdimen\pointwidth 
\setbox0=\hbox{\rm .} 
\pointwidth=\wd0 
\catcode`?=\active 
\def?{\kern\pointwidth} 
\halign{\hbox to 1.0in{#\leaderfil}\tabskip=2.0em&
\hfil#\hfil\tabskip=1.0em&
\hfil#\hfil\tabskip=1.0em\cr
\noalign{\doubleline}
\omit\hfil Parameter\hfil& \multispan2\hfil $1\sigma$ values\hfil \cr
\omit & correct method& incorrect method\cr
\noalign{\vskip 3pt\hrule\vskip 4pt}
$\ln (\alpha_{\rm em})$  & 0.0058@   & 0.0056@ \cr
$\ln (\alpha_{\rm g})$   & 0.031@@   & 0.062@@ \cr
$\ln (\theta_{eq})$      & 0.0096@   & 0.04*@@ \cr
$\ln (\theta_0)$         & 0.0016@  & 0.025*@ \cr
$X$                & 0.10@@@   & 0.17*@@ \cr
$A_{\rm s} \times 10^{9}$        & 0.11@@@   & 0.105@@ \cr
$n$                & 0.0074@   & 0.0073@ \cr
$\tau$             & 0.021@@   & 0.021@@ \cr
$h_0$              & $1.67^*$@@ & 1.38@@@ \cr
$\Omega_{\rm B}$   & $0.0017^*$ & 0.00091 \cr
$\Omega_{\rm CDM}$ & $0.0056^*$ & 0.0032@ \cr
\noalign{\vskip 3pt\hrule\vskip 4pt}
}}
\endtable
\caption{Comparison of the $1\sigma$ error bars of the cosmological
parameters of the two different methods described in Section~\ref{sec:fisher} 
for the Fisher matrix analysis.  The first approach involves the
cosmological model in an explicitly dimensionless way, while in the second
approach $G$ is simply added as a parameter to the usual model. A star on a value
shows that the parameter was not among the set of parameters for Fisher matrix
calculation in the relevant method. }
\label{tab:fisher}
\end{table*}

\section{Other cosmological observables}
We have already considered the effects of a variable $\alpha_{\rm g}$ or
$\alpha_{\rm em}$ on BBN and CMB anisotropies, with the main emphasis on
making things dimensionless.  One can extend this approach to other
observables in cosmology, such as weak lensing, baryon acoustic oscillations
(BAO), large-scale structure, ISW effect
and the use of supernovae as standard candles.  We will avoid discussing the
use of high redshift quasar spectroscopy to constrain $\alpha_{\rm em}$
and $\mu$, since that topic is extensively covered elsewhere (see references
in \cite{Uzan2011}).

\subsection{Baryon acoustic oscillations}
BAO follows the same basic physics as CMB anisotropies (see e.g.\
\cite{EisensteinHu}), so there should in
principle be the same kind of $\alpha_{\rm g}$ dependence.  The important
timescales for BAO are the equality epoch, $z_{\rm eq}$ and the drag epoch,
$z_{\rm d}$.  The first, $z_{\rm eq}$, is purely determined by the initial
conditions of the Universe, and the time today, setting the epoch when
$\Omega_{\rm M}=\Omega_{\rm R}$.  The drag epoch is defined as the time when
baryons are freed from the Compton scattering of photons, and therefore
$z_{\rm d}$ has the same $\alpha_{\rm g}$ or $\alpha_{\rm em}$ dependence as
$z_{\rm r}$.  These special redshifts set the scaling conditions for BAO.
After the drag epoch, one should solve for the matter transfer function,
where this function is basically the same as the CMB transfer function in
terms of $\alpha_{\rm g}$ or $\alpha_{\rm em}$ dependence.
It seems clear therefore, that if one wanted to use BAO to constrain the
variation of $\alpha_{\rm g}$, it would be straightforward to follow the
same procedure we discussed in section~5.

\subsection{Gravitational Lensing}
In gravitational lensing (see \cite{KraussWhite} for an earlier study)
there is a danger of becoming confused by dimensions, since both the estimate
of the lensing mass and the curvature depend on $G$, and
they are mixed together in $\alpha_{\rm g}$.  The simple case of an
Einstein ring is illuminating.  Here the lens and the source are colinear
and in Euclidean space the radius of the Einstein ring is the geometric
mean of the Schwarzschild radius of the lens and the distance to the lens.
This means that for a relatively nearby lens, at distance $d$, the lensing angle
is $\Theta_{\rm E} \sim \sqrt{G M/c^2 d}$, and if we estimate the mass through
measuring a velocity dispersion $v^2$ over a radius $r$, then
$\Theta_{\rm E} \sim \sqrt{\Theta v^2/c^2}$, where $\Theta\equiv r/d$ is the
apparent angular size of the object.  Viewed this way we see that it is hard
to use lensing to measure the strength of gravity, because the obvious
dimensionless observables leave no $G$ dependence!

Let us look at this in a little more detail.  In cosmology
one can work out the lensing angle to be (e.g.\ \cite{sean})
\beq
{
\Theta_{\rm E} = \sqrt{\frac{4 G M d_{\rm LS}}{c^2 \, d_{\rm L} \, d_{\rm S}} },
}
\eeq
where $d_{\rm LS}$ is the distance from the source to the lens, and $d_{\rm S}$
and $d_{\rm L}$ are, respectively, the distances from us to the source and to
the lens.  However, this seemingly innocent dimensionless equation is not
useful for any experiment designed to measure a change in the gravitational
constant; this equation, or any equation for lensing, should be turned into an
equation which only has $\alpha_{\rm g}$ or $\alpha_{\rm em}$ dependence
before it can be used as a test of fundamental constant variation.  One can
work out all of the distances in the above equation and end up with the
following relation:
\begin{eqnarray}
\Theta_{\rm E} &=& A \,\alpha_{\rm g}^{3/4} \,\theta
 \,\dfrac{M}{m_{\rm p}}, \\*
{\rm with\ } A & \equiv & \sqrt{\dfrac{8 \, \pi^3}{45 \Omega_{\rm R}}} \, 
 \sqrt{\dfrac{1+z_{\rm L}}{N_{\rm L}}-\dfrac{1+z_{\rm L}}{N_{\rm L}}}.
\end{eqnarray}
Here $z_{\rm L}$ and $z_{\rm S}$ are, respectively, the redshift of the lens
and the source and $N_{\rm L}$ and $N_{\rm S}$ are defined as:
\beq
{
N_x \equiv \int_{\dfrac{1}{1+z_x}}^1 \, 
\dfrac{dy}{y^2\, \sqrt{\Omega_{\rm \Lambda} \, + \, \Omega_{\rm M} y^3 \, 
\Omega_{\rm R}y^4}},
}
\eeq
where $x$ can be either S or L.

Speaking in more general terms (and again thinking about communicating
with another civilization), we can see two possibilities.  The first is that
one could imagine choosing a lens which consists of a fixed number of
particles, so that one knows the value of the pure number $M/m_{\rm p}$.
That may be interesting from a philosophical point of view,
but is hardly related to how we carry out observations in cosmology.
A second idea is that perhaps one could perform a statistical survey over
many galaxy lenses, measuring statistics which depend on a characteristic
galaxy mass $M_{\rm gal}$.  If that mass depends on fundamental constants
through a cooling argument (e.g.\ \cite{ReesOstriker}), then it may be that
in a lensing survey the observables depend on the number
$\alpha_{\rm em}^5\alpha_{\rm g}^{-2}\mu^{-1/2}$.  Since this would be
different in universes with different values of the constants, then this
is potentially measurable, and hence it {\it might\/} be possible to
constrain a redshift dependence of lensing observables.

These ideas do not seem particularly practical, and so we are left unclear
about whether gravitational lensing could ever be used to constrain the time
variation of $\alpha_{\rm g}$.  Things will presumably be unambiguous in
explicit self-consistent models which contain a variable strength of gravity,
and we leave this topic for future studies of specific models.

\subsection{Large-scale structure and supernovae}

There are many other astrophysical phenomena which can in principle be used
to test the variation of fundamental constants (see e.g.\ \cite{Uzan,GBIK}).
Large-scale structure and various measures of the power spectrum of galaxy
and matter clustering \cite{Nesseris2011},
will also have dependence on $\alpha_{\rm g}$.  It
will again be important to ensure that this dependence is dimensionless,
and that cosmological parameters are not over-constrained when doing so.
In other words, one needs to realise that changing $\alpha_{\rm g}$ can
effectively change the epoch at which we live.

Supernovae have also been very useful as approximately standard candles
in cosmology, and such studies can also be adapted to constrain a
combination of fundamental constant variations
\cite{Gaztanaga,GBetal2006,NesserisP,DunganP}.  Like other stellar sources,
there is a dependence on $\alpha_{\rm g}^{-1/2}$ in the evolutionary
timescale (see e.g.\ \cite{Maeder,Innocenti,Thorsett,Guenther98,GBetal2011}),
and hence the use of supernovae, combining the standard candle
property with luminosity distance, will involve a different combination of
fundamental constant variations than for BAO and lensing studies.
A combination of cosmological probes will therefore enable variations among
different parameters to be distinguished.

Finally, we note that among the different arenas within astrophysics,
cosmological studies have the potential of probing not only the time
dependency of $\alpha_{\rm g}$, but also any scale dependence.
It is important to work out an appropriately consistent theory to describe
these phenomena within the context of a spatially variable fundamental
constant.  Although we have not considered such models here,
it will surely still be important to avoid dimensional quantities.

\section{Conclusions}

We are not advocating that all of cosmology henceforth should be represented
in dimensionless forms.  However, we would say that care has to be taken
when dealing with variation of physical constants in cosmology.
This is particularly because of the cosmic time ambiguity, which could lead
to {\it over}-constraining the cosmological parameters, especially when
$G$ is involved.

Definite physical mechanisms for the variation of fundamental constants
will lead to specific forms for $\alpha_{\rm g}(t)$ etc.  We believe that
the proper place to start investigating such theories is to make sure that
there is a robust basis for comparing cosmologies with different
{\it constant\/} values of the constants.  Only then can one effectively
deal with time-variable quantities.

A time-dependent (or space-dependent) $\alpha_{\rm g}$ is not consistent with
General Relativity.  One explicit framework for accommodating such a variation
are the scalar-tensor theories of gravity.  Some studies have already
investigated CMB anisotropies and other cosmological constraints in the
simplest form of scalar-tensor model (e.g.\ \cite{kamion,acquaviva,chen}),
the so-called Brans-Dicke theory \cite{dicke2}.  While this version of a
scalar-tensor theory seems to be already ruled out by solar system
experiments \cite{rea}, there are more general scalar-tensor theories which
pass the solar system tests \cite{nordt}, and might be promising avenues of
exploration (e.g.\ \cite{dirac1,Tsujikawa}).
These may provide analogues for investigating the variation
in physical constants proposed in some brane-world scenarios.
In studying the empirical tests of such models we recommend keeping
everything dimensionless in order not to be misled by apparent variations
that may be unmeasurable.

\section*{Acknowledgments}
This work was supported by the Natural Sciences and Engineering Council of
Canada and by the Canadian Space Agency.

\bibliographystyle{unsrt}
\bibliography{constant2}

\end{document}